# The Landscape of the Quantum Start-up Ecosystem

Zeki Can Seskir[1]*, Ramis Korkmaz[2], Arsev Umur Aydinoglu[3]

[1] Institute for Technology Assessment and Systems Analysis, KIT, Karlstraße 11, 76133 Karlsruhe, Germany
[2] Department of Economics, METU, Üniversiteler, 06800 Çankaya/Ankara, Turkey
[3] Science and Technology Policy Studies, METU, Üniversiteler, 06800 Çankaya/Ankara, Turkey

* zeki.seskir@kit.edu

**Abstract**

The second quantum revolution has been producing groundbreaking scientific and technological outputs since the early 2000s; however, the scientific literature on the impact of this revolution on the industry, specifically on start-ups, is limited. In this paper, we present a landscaping study with a gathered dataset of 441 companies from 42 countries that we identify as quantum start-ups, meaning that they mainly focus on quantum technologies (QT) as their primary priority business. We answer the following questions: (1) What are the temporal and geographical distributions of the quantum start-ups? (2) How can we categorize them, and how are these categories populated? (3) Are there any patterns that we can derive from empirical data on trends? We found that more than 92% of these companies have been founded within the last 10 years, and more than 50% of them are located in the US, the UK, and Canada. We categorized the QT start-ups into six fields: (i) complementary technologies, (ii) quantum computing (hardware), (iii) quantum computing (software/application/simulation), (iv) quantum cryptography/communication, (v) quantum sensing and metrology, and (vi) supporting companies, and analyzed the population of each field both for countries, and temporally. Finally, we argue that low levels of quantum start-up activity in a country might be an indicator of a national initiative to be adopted afterwards, which later sees both an increase in the number of start-ups, and a diversification of activity in different QT fields.

**Introduction**
For over the last two decades, the second quantum revolution [1] has been ongoing with increasing global interest. Since the early 2010s, many of the large economies such as the UK [2], the US [3], France [4], Germany [5], Russia [6], India [7], and China [8] initiated national programs to support the development of quantum technologies (QT). In addition to the national programs, regional initiatives such as the EU Quantum Flagship [9], and international collaboration within greater contexts, such as the promise of increased collaboration on QT under AUKUS [10], have been happening in the last few years. These are being accompanied by a sharp increase in both the number of academic publications [11], the number of patents granted [12], and overall investments [13], which have led to an increased amount of private funding in a period some refer to as the 'quantum gold rush' [14]. The expected public investment to QT within the next five to ten years is more than 20 billion euros [15]. In just a decade, QT became a contender for being the 'next big thing' from a topic discussed mainly in physics conferences.

All these activities are also accompanied by an increase in the number of start-ups being founded, and an ecosystem rapidly growing. Some aspects of the QT ecosystem in general have been studied in the literature, such as the skills and knowledge needed for the required new workforce [16], and the temporal consultancy work within the growing QT market [17]. Furthermore, there are many studies focusing on applications of QT and its subfields, such as industry [18] and commercial [19] applications of quantum computing, quantum computing for finance [20,21], chemistry [22], and so on. Similarly, there are many studies on sectoral applications of QT like for space [23], telecommunications [24], and military [25]. However, a study on the emerging landscape of the quantum start-up ecosystem has been missing in the literature.

In this paper, we aim to answer several essential questions regarding the evolution and properties of this emerging landscape:
(1) What are the temporal and geographical distributions of the quantum start-ups?
(2) How can quantum start-ups be categorized, and how are these categories populated?
(3) What type of patterns can be derived from empirical data on trends?

**Methods**
In this section, we provide information on how the dataset of quantum start-up companies were compiled, how they were categorized under six subfields, and the limitations of our methods.

*Dataset*
The dataset utilized in this study was created between September 2017 and September 2022 with gradual additions. We mainly used publicly available resources such as websites, LinkedIn profiles, targeted Crunchbase searches [26], and other databases. There were several ordered lists of companies that we utilized as resources, such as the lists in at Understanding Quantum Technologies document by Olivier Ezratty [27], the Private/Startup Companies section of the Quantum Computing Report website [28], The Quantum Insider platform [29], QIS Data Portal [30], and several consortium member lists such as Quantum Industry Canada (QIC) [31], The Quantum Economic Development Consortium (QED-C) [32], European Quantum Industry Consortium (QuIC) [33], UKQuantum [34], Quantum Technologies Development Consortium (QTC) in Israel [35], Quantum Technology and Application Consortium (QUTAC) in Germany [36], MSU Quantum Technology Centre in Russia [37], Quantum Business Network in Germany [38], and IBM Quantum Network [39]. Furthermore, in years we added start-ups to our database from the participant lists of events like Inside Quantum Technology [40], Quantum Business Europe [41], Q2B [42] organized by QCWare, and Careers in Quantum Technologies organized by QURECA [43].

First of all, we wanted to correctly depict the evolution of the dedicated quantum start-up ecosystem; hence, we aimed at excluding all companies that only approached QT as a secondary business or focused on it only later in their company history. Our aim was to minimize the number of false positives in our database. It was easy to identify and exclude major corporate players such as IBM and Google. It was also easy to identify many of the start-up companies that were operating in adjacent fields, such as photonics, nanotechnology or cybersecurity, and getting into the quantum field. There were some borderline cases, start-up companies that started in an adjacent field but transitioned to prioritizing QT in recent years. To identify these, we used the WayBackMachine provided by the non-profit Internet Archive [44] to check previous versions of the company websites, and look for explicit references to QT or closely related products (like single photon sources-detectors, cryogenic systems, quantum dots, etc.).

Secondly, we wanted to identify the countries and founding years of these start-ups as accurately as possible. This especially became a priority after encountering numerous contradicting information on publicly available databases. Therefore, we adopted the following double-check mechanisms: (i) checking when the website was founded via WayBackMachine, (ii) checking LinkedIn profiles of one or several founders, (iii) checking news articles (especially useful for university spin-offs), and (iv) checking seed or series A funding news if there are supporting venture capital firms. Sometimes more informal methods, such as asking an employee or a founder the founding date of the company, were also employed.

Thirdly, for the country of the start-ups, if the company moved after its founding, we considered the original country as the 'Country' of our start-up. This was a deliberate choice to avoid any bias that might be carried on by the date data gathering was stopped. Also, we believe that a company's origin country has more explanatory value in identifying the global distribution of expertise and interest, and a company's operating country has more explanatory value in depicting the current market maturity levels. Researchers interested in market activity are welcome to utilize and alter our dataset with a quick check on each start-up.

Finally, during this paper we refer to all the new companies that were founded with a particular focus in the QT as quantum start-ups, regardless of their age. This is due to our analysis mainly focusing on the founding of these companies, rather than their performance, longevity, or other qualities. For the dataset, we acknowledge that there are some companies that shut down their operations, and others that were acquired by larger companies, and a few that went through mergers. We did not incorporate this information into our dataset, again for the same reason above. Formation of a start-up is an indicator of at least some expertise and interest in that country. Similarly, shutting down operations, acquisitions, and mergers are markers for different types of market activities, and although these might bear insightful fruits for innovation studies purposes, they are beyond the scope of our study.

*Encoding*
After compiling the dataset, we went through the process of assigning a main field of operation to each start-up. Following the US framework [3] we divided QT into three main subfields quantum computing/simulation, quantum cryptography/communication, and quantum sensing and metrology. After this, due to the levels of high activity in quantum computing, we divided it into two subfields of quantum computing (hardware) and quantum computing (software/application/simulation). Additionally, we added two more fields to cover companies that could not be listed under these previously mentioned four fields, and those were complementary technologies, and supporting companies. So, in the end, we had the following six fields: (i) complementary technologies (sometimes also referred to as enabling technologies), (ii) quantum computing (hardware), (iii) quantum computing (software/application/simulation), (iv) quantum cryptography/communication, (v) quantum sensing and metrology, and (vi) supporting companies.

Complementary technologies cover start-ups developing cryogenic systems, control electronics, vapor cells, and similar products that particularly aim to enable QT devices. Quantum computing hardware companies are those that focus on developing the hardware of quantum computers (usually qubit technologies) relying on different types of physical systems such as ion traps, superconducting systems, photonics, silicon, diamonds, and others. Quantum computing software/application/simulation field includes companies from those developing sector-specific applications (like finance) to companies working on quantum software development (such as compilers). For the quantum cryptography/communication field, we decided to include dedicated post-quantum cryptography (PQC) companies, in addition to the expected quantum key distribution (QKD), quantum encryption, and quantum network companies. The quantum sensing and metrology field includes both types of companies that are developing quantum sensors and the ones that utilize quantum sensors for sector specific solutions. Finally, supporting companies are start-ups that exist solely for purposes like consulting, education, event organization, and incubators.

For visualization, we used *ggplot2*, *ggforce*, *RColorBrewer*, *dplyr*, and *forcats* packages in the R programming language. We used color blind friendly palettes when possible.

*Limitations*
There are three main limitations of this study. First, although we believe our dataset to be highly representative, we cannot argue that it is exhaustive. There are start-ups with no online activity. This is particularly true for non-English speaking countries. Therefore, it should be noted that the database biases Western countries, especially the English-speaking ones. Secondly, the exclusion principles upheld here can easily be challenged. One might argue that a start-up that now focuses on QT should be accepted as a quantum start-up, which is as valid as our claim. We aimed at keeping the false positives near zero, which might have yielded an increased number of false negatives, and caused the exclusion of legitimate quantum start-ups. Hence, this limitation of arbitrariness in where the line is drawn should also be acknowledged while considering our outcomes and analysis. Finally, although here we focused on the number of start-ups to describe the landscape of the quantum start-up ecosystem, investment, valuation sizes, and similar parameters can also be included as additional layers of analysis. Therefore, one should keep in mind the limitations of this study while assessing its conclusions.

In the next section, we present the key properties of our dataset and the outcomes of our analyses.

**Results**
Our dataset covers 441 companies in 42 countries that mainly focus on QT as their primary priority business. The oldest companies covered in our dataset are D-Wave Systems and MagiQ Technologies, which were founded in 1999. Therefore, the dataset covers 23 years of data between 1999-2022. We did not include corporate divisions (like IBM Quantum or Google Quantum AI Lab) in our dataset, and focused on and included companies that were founded as start-ups in QT.

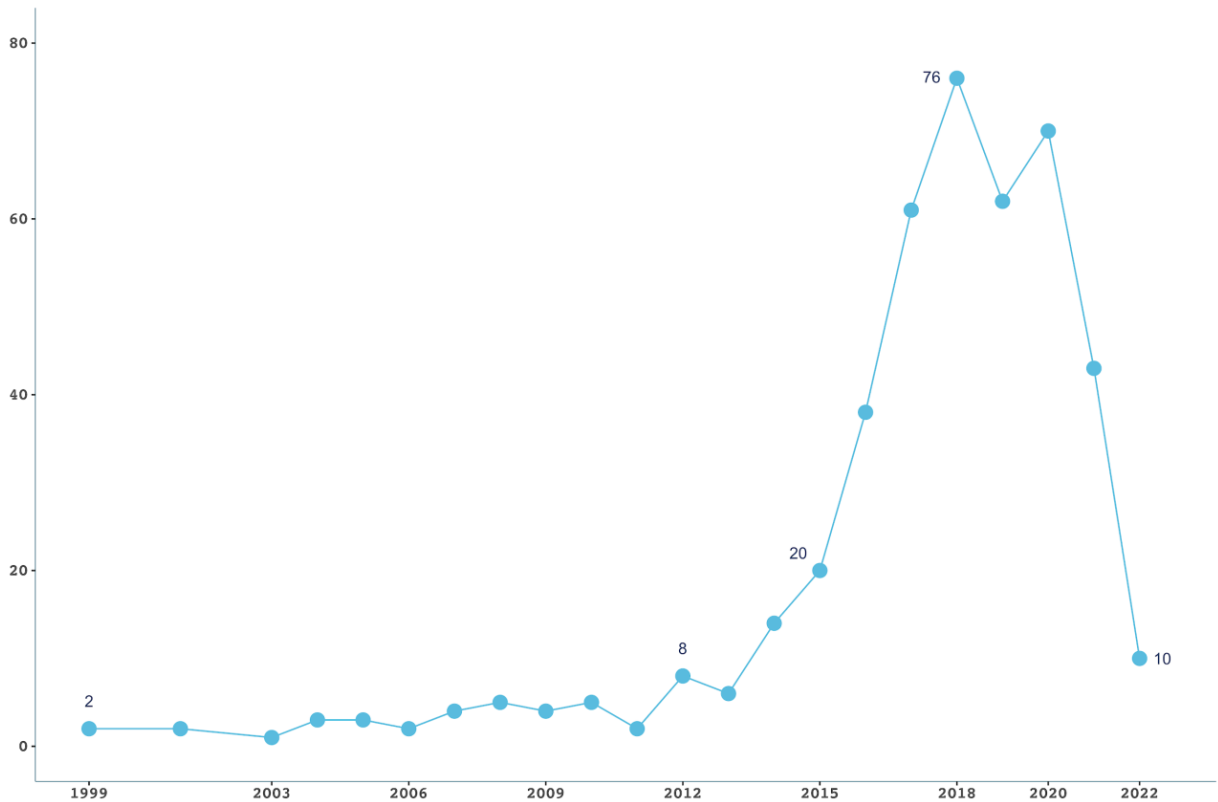

Figure 1: Number of new quantum start-ups per year.

In Fig. 1, the yearly distribution of new start-ups per year is given. In our dataset, it appears that the number of start-ups founded has increased sharply starting from 2013, peaked in 2018, and has been slightly declining ever since (Fig. 1). In Fig. 2, the global distributions of these companies are provided. Though it might be important to keep in mind that our dataset does not cover stealth start-ups with zero visibility and has a lower representation of start-ups from non-English speaking countries. This can also be observed in the global distribution of start-ups (Fig. 2), as the majority of companies in the dataset are from the US, Canada, and the UK.

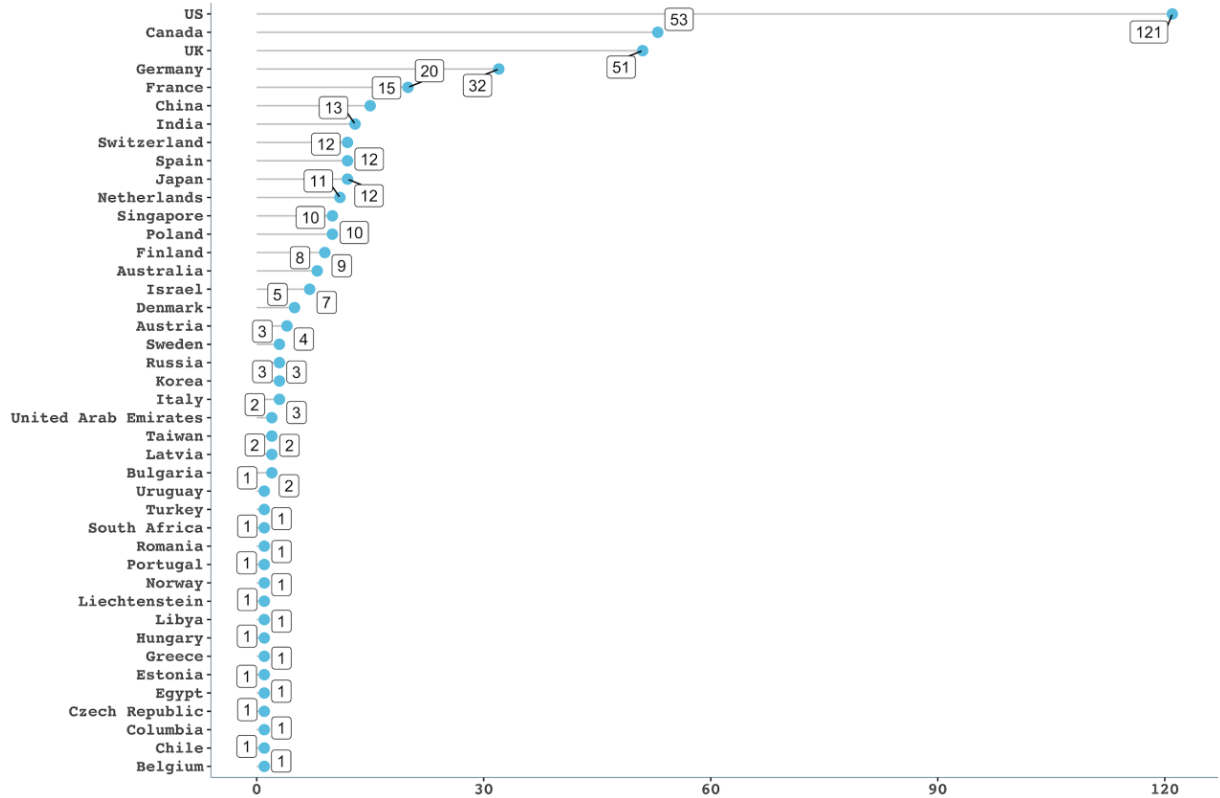

Figure 2: Number of quantum start-ups in countries.

Next, we checked the yearly distribution of new start-ups with respect to different fields. In Fig. 3, the violin plot shows the temporal distribution of start-ups in fields. The areas inside the curves add up to the total number of start-ups in respective fields.

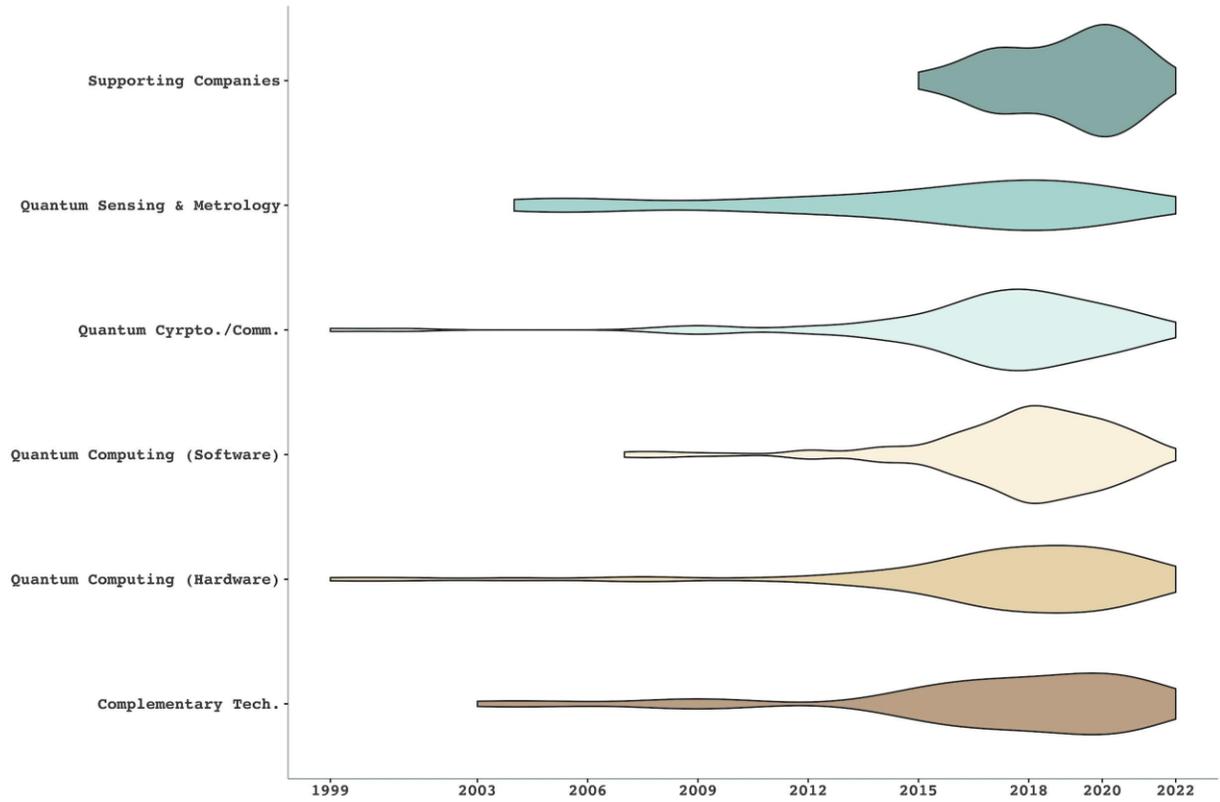

Figure 3: Violin plot of new quantum start-ups per year in different fields.

It can be seen from Fig. 3 that almost all fields have experienced a surge of interest in recent years, although there has been some activity in the field since the early 2000s. The most uniform temporal distribution can be seen in the quantum sensing and metrology field, while the most skewed distribution is for the supporting companies. This can also be observed in Fig. 4, where we looked at numbers per year in different fields instead of ratios.

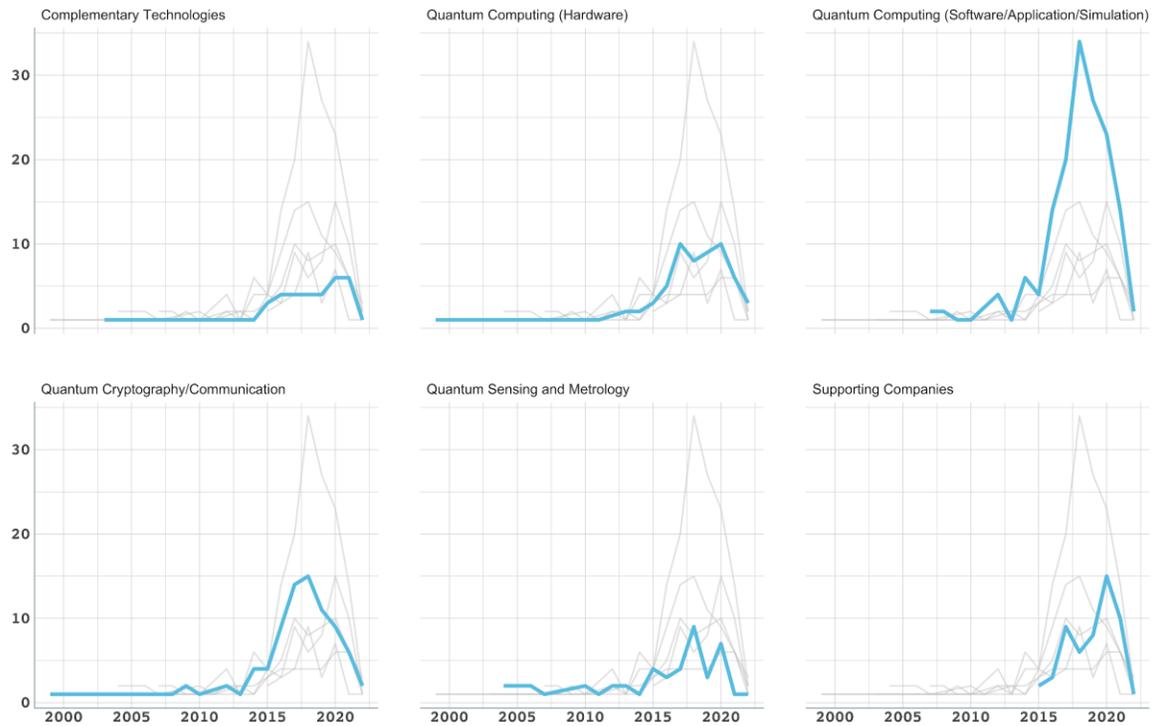

Figure 4: Number of new quantum start-ups in their respective fields per year.

Following this, we checked the global distribution of start-ups in different fields in Fig. 5. It can be seen that the most diverse fields (where a high number of countries are active) are quantum computing (software/applications/simulation), quantum cryptography/communication, and supporting companies. The least diverse fields are quantum sensing and metrology, and complementary technologies. Although, regardless of diversity, the US leads in the number of start-ups in every field, while the second place varies between fields. For example, Canada has the second most start-ups in both quantum computing fields (hardware and software/applications/simulation), while the UK has the second place in the quantum sensing and metrology, and the quantum cryptography/communication fields. Lastly, in the complementary technologies field, Germany and Netherlands are competing for second place.

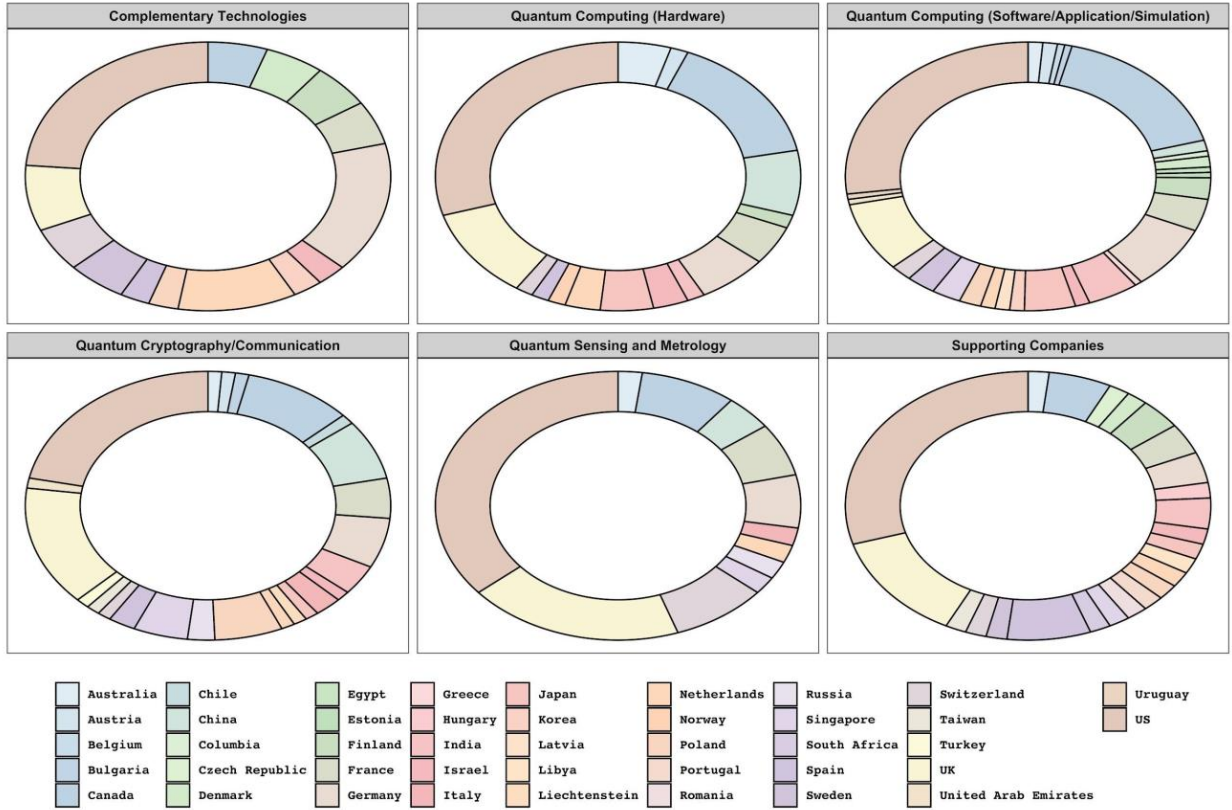

Figure 5: Pie chart of each field with ratios of countries represented in different colors.

Afterwards, we checked the global distribution of start-ups in different fields in Fig. 6. For the top five countries (the US, Canada, the UK, Germany, and France), it can be seen that each six fields are represented. However, as we go down the line, it becomes less frequent. We can see that in some countries, only supporting companies exist. Furthermore, there is no apparent pattern across the countries in the ratios of start-ups operating in different fields.

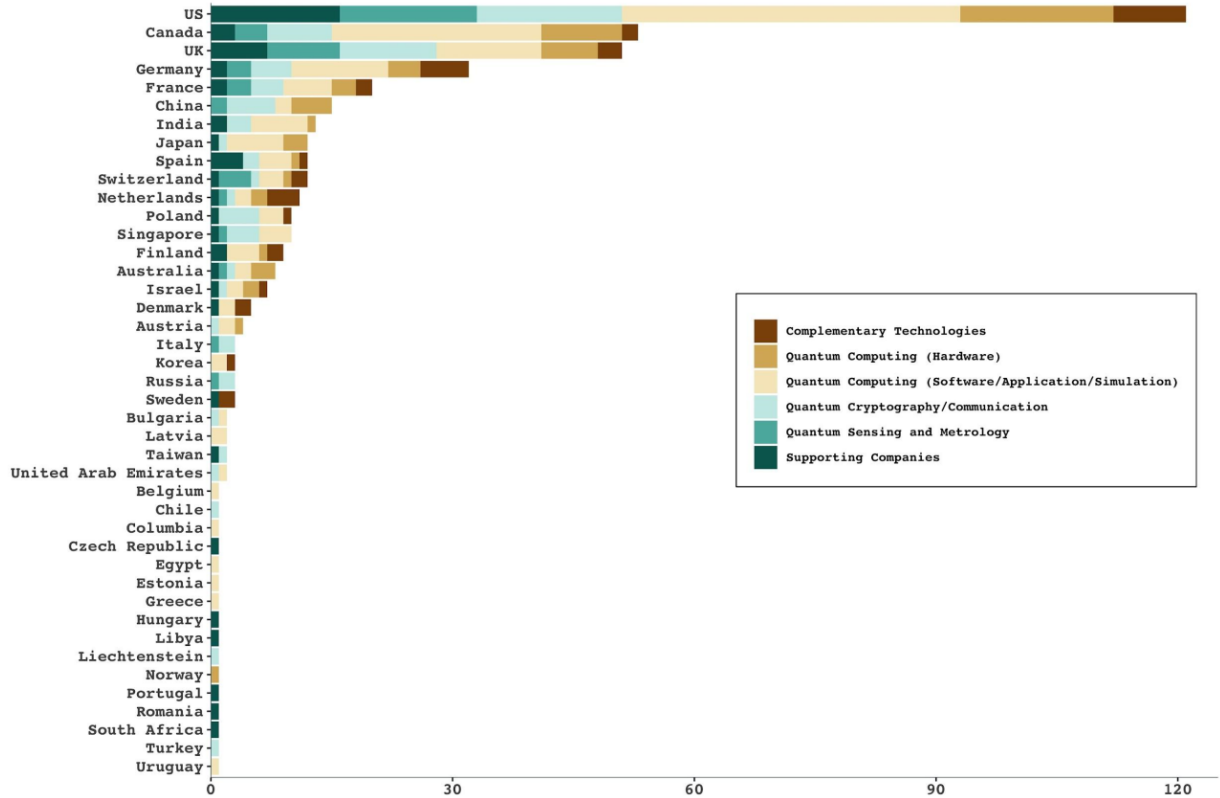

Figure 6: Number of quantum start-ups in countries in different fields stacked.

In Fig. 7 we present the information on which and when fields have been emerging in which countries. It can be seen that in the US and the UK, supporting companies have been the emerging trend for the last several years, while in other countries, the trends vary widely. This is particularly useful to see in countries with a low number of start-ups, which field was introduced when. It can be seen that almost half of the countries in the list had their first start-up in quantum technologies following 2015, and the level of activity in countries with already one or few start-ups has highly increased since 2015. However, it can be seen that in a handful of countries, there has been activity since the mid -2000s, and in the US, there has been almost continuous activity since the early 2000s.

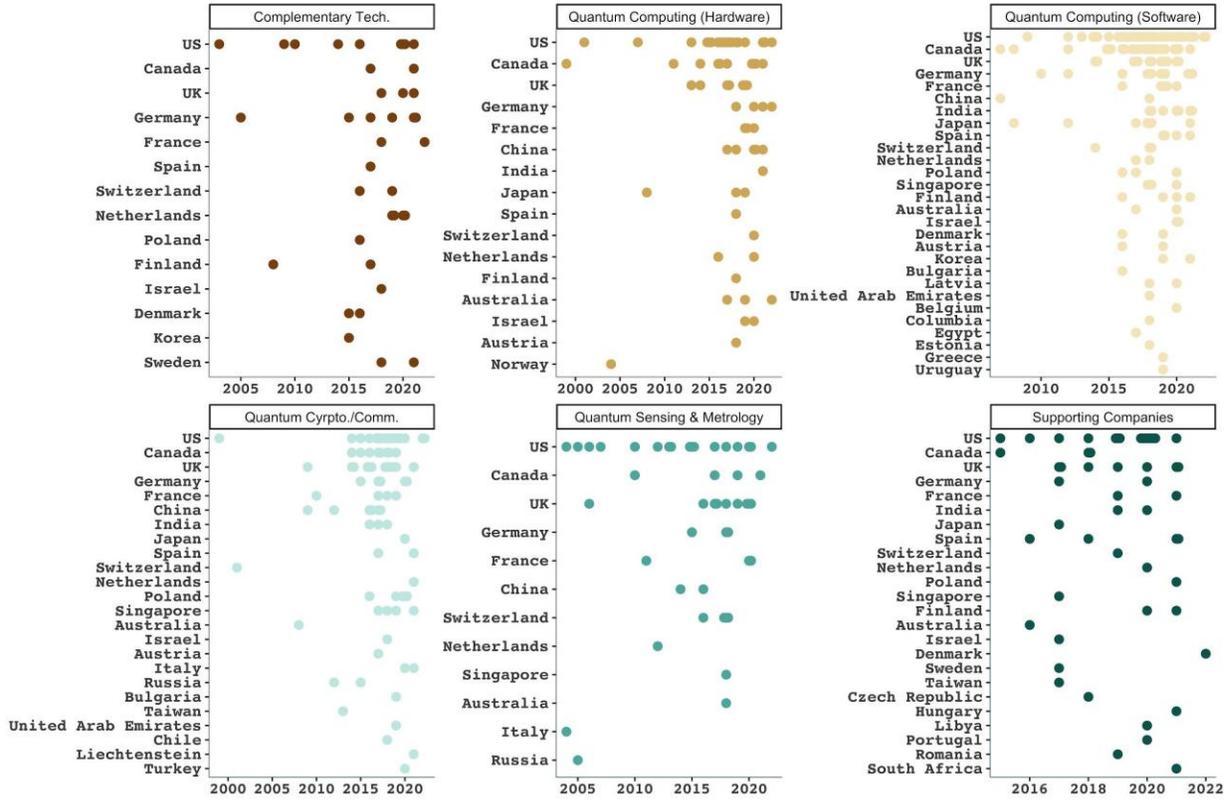

Figure 7: Yearly frequency of companies founded for each country to identify emerging interests in fields.

Following the country based analysis, we clustered the start-ups founded in different regions for a global comparison of regions versus countries. This was particularly conducted to compare the EU-27 to the US, Australia, Canada and the UK (acronymed as AU-CA-UK), and the rest of the world (acronymed as ROTW). Two figures were created for comparison. In Fig. 8, we compare the total number of companies founded in these regions, and in Fig. 9, different fields in a stacked graph.

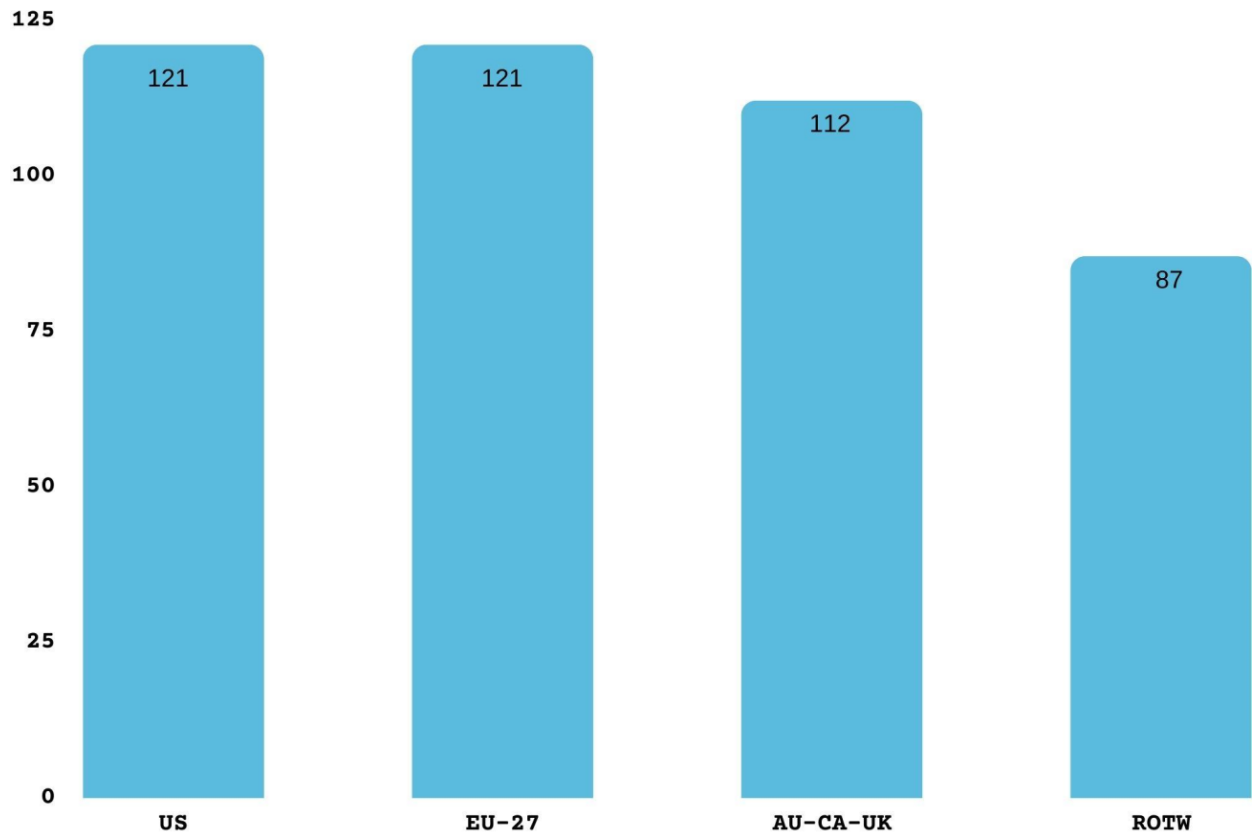

Figure 8: Total number of quantum start-ups founded in different regions.

Fig. 8 demonstrates the additional insights that can be gained by comparing regions instead of countries. Here, it can be seen that in terms of sheer numbers of companies the US, the EU-27, and the AU-CA-UK are operating on similar levels. Meanwhile, the rest of the world does not possess that many start-ups. This can be related to the limitations of our dataset, the fact that start-up culture is more adopted by certain regions, some other factors or a combination of these.

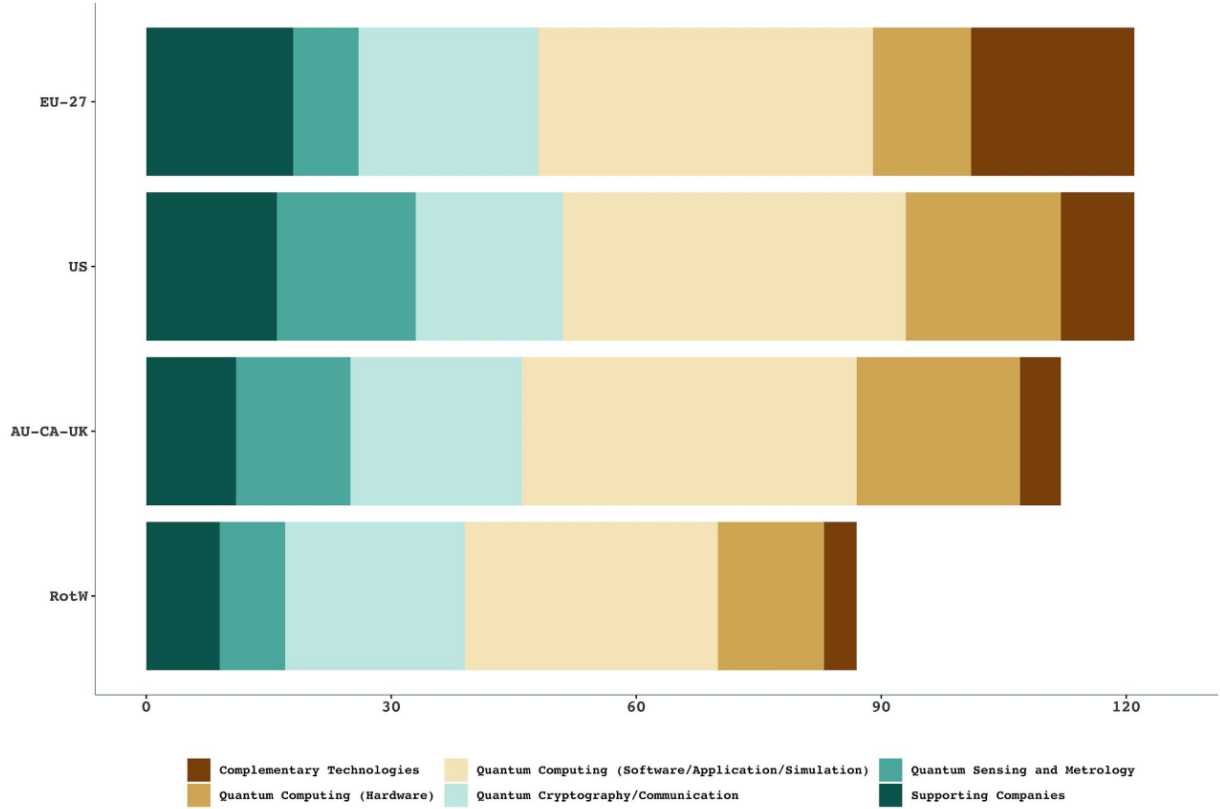

Figure 9: Number of quantum start-ups in regions in different fields stacked.

Finally, in Fig. 9, we present the type of companies in different regions. It can be seen that the EU-27 has more supporting companies, start-ups focusing on complementary/enabling technologies and quantum cryptography/communication start-ups, while the US has a strong lead compared to the EU in quantum sensing and metrology, and quantum computing hardware start-ups. This figure shows that the adoption of different fields of QT varies on regional levels, however, all regions have presence in all the fields of QT.

**Discussion**

There are several quick answers that we can derive from the data presented above to our research questions. First of all, the temporal distribution of the quantum start-up ecosystem is not uniform; more than 92% of the companies we identified as quantum start-ups have been founded within the last 10 years, 68% in the last 5 years. Similarly, for the geographical distribution, more than 50% of the start-ups are located in the US, the UK, and Canada. Although there are 42 countries in our dataset, only half of them have more than two start-ups. We can say that the phenomenon of the quantum start-up ecosystem is both recent and limited in its global adoption.

Secondly, we categorized the companies under six fields, and investigated how these fields are populated. It can be seen in Fig. 3 that the most uniformly populated field is quantum sensing. Furthermore, it can be argued that there are established companies that are filling the market gap that might provide the necessary incentive for a start-up to be founded. We observed this while gathering data, as several measurement device manufacturers adopted QT-oriented products as one of their primary businesses, and we had to go back in their corporate history to identify whether they were quantum start-ups from the beginning; if not, we excluded them from the dataset. In Fig. 3 and Fig. 4 it can be seen that all fields were most active in the period after 2016, which again is not surprising since both the EU Flagship [9] and the US National Quantum Initiative [3] were kicked off in 2018, signaling that there will be increased market activity and funding opportunities in the following years.

Thirdly, a phenomenon to be observed regarding the categories is the emergence of a field of 'Supporting Companies'. The consultancy aspect of this phenomenon was also addressed previously in the literature for quantum computing in a qualitative manner [17}. Here, our data clearly shows the emergence of supporting companies around the newly established QT sector. The businesses of these companies are mainly consultancy, ecosystem support/development, reporting, event organization, and education. As the needs of the quantum industry [16] diversify, it is only natural that the supporting ecosystem becomes diverse in the services they provide to this rapidly growing sector.

Fourthly, there are two clear patterns from the gathered data that we observed. First, the number of countries that contain interested entrepreneurs toward QT has been increasing, especially since 2016. Additionally, it appears that low levels of start-up activity usually precede the formulation of national initiatives, and the number of start-ups rises after such initiatives are announced. In the light of this observation, it can be proposed that for a country to formulate a national initiative, having one or two start-ups in the QT domain is usually a precursor or a catalyst. Furthermore, if this can be generalized, we can expect national initiatives in more than 20 countries where start-ups emerged following 2016, and that does not have a national initiative focusing on QT. Secondly, 38% of countries possess QT start-ups that are operating in four or more fields of QT, indicating a diverse set of activities in QT. This by itself is not an indicator that different fields in QT are supporting each other, but remembering from Fig. 5 and Fig. 7 that almost all fields have certain levels of globally diverse start-up actors. To summarize this point, low levels of activity might be an indicator of a national initiative for a country, which later sees both an increase in the number of start-ups, and a diversification of activity in different QT fields.

Fifthly, it should be noted that not all start-ups are created equal. For example, PsiQuantum, a photonic quantum computing hardware manufacturing company, has raised over 600 M$ in its funding rounds [45] or IonQ, a publicly-traded ion trap quantum computing hardware manufacturing company, is valued almost around 2 B$ [46]. On the other hand, many of the start-ups have much lower seed funds or initial investment amounts [47].

Sixthly, an open question in the QT ecosystem development predictions is whether there is room for further diversification or if the time for consolidation is soon. We recently started seeing mergers in the field, such as Honeywell Quantum Solutions and Cambridge Quantum [48], and Pasqal and Qu&Co [49]. Also, there have been partial acquisitions, such as SK Telecom buying out the majority share of ID Quantique [50]. There have also been special purpose acquisition company (SPAC) deals like IonQ [46], Rigetti [51], and D-Wave [52], which attracted hundreds of millions of dollars each, but also potentially created new hurdles for the companies such as more public scrutiny and hostile reports by short-sellers [53]. As discussed in the literature previously, the quantum gold rush (14) caused the attraction of private funding to the field in many shapes and forms. However, the phenomenon of acquisitions, mergers, and SPACs appear to be recent, which followed the increased interest in the QT domain following 2016. Additionally, there are new mechanisms to support diversification, such as the European Innovation Council (EIC). Previously, EIC supported three companies in the QT field [54] and for 2022 there is an open Pathfinder Challenge in "Alternative approaches to Quantum Information Processing, Communication, and Sensing" with an approximately €28M [55] budget. Therefore, the question of whether the trend of diversification will continue at this rate or whether consolidation will become the modus operandi of the field is a question directly related to whether further increase in interest from countries and private investment sources are possible or not. The idea of a possible quantum winter, especially in quantum computing [56], has been discussed previously. Our data does not clearly provide any answers in one way or another; however, running a similar analysis in two or three years and comparing with ours might allow further insight, especially on the question of whether 2018 was a true maximum, or that we are just missing data due to the limitations listed in the Methods section.

Seventhly, we expect a further expansion of categories will become necessary as the field progresses. Within the field of quantum cryptography/communications, there are companies that are engaged in PQC, QKD, quantum encryption, and quantum networks. The timelines of these subfields vary, hence, as the number of start-ups and market activity increases, this field is probably going to require a division into three subfields; PQC, quantum cryptography, and quantum communication. A similar argument can also be made for the quantum sensing and metrology field, since it encompasses companies operating in increasingly separate areas of activity.

Eightly, it is clear from Fig. 8 and Fig. 9 that regional level analysis offers additional insight on top of the country level analyses. However, reaching conclusions from regional level analysis require further investment into identifying the appropriate regional clusters. In this work, we focused on comparing the EU-27 with the US and AU-CA-UK as potential market competitors, but the relationship between these regions and the clustered countries deserve a dedicated analysis. Our preliminary findings here reflect that there are regional differences in preferences towards certain fields of QT, but there is diversity of companies in terms of fields within each region.

Finally, while gathering data, we noticed the emergence of a new group of organizations within the quantum ecosystem, which are non-profit organizations and similar supporting entities. There are several consortia of organizations mainly aiming at industry goals; Quantum Industry Canada (QIC) [31], The Quantum Economic Development (QED-C) in the US [32], European Quantum Industry Consortium (QuIC) [33], Quantum Technology and Application Consortium (QUTAC) in Germany [36], UKQuantum [34], Quantum Ecosystems Technology Council of India (QETCI) [57], Quantum Technologies Development Consortium (QTC) in Israel [35], Quantum Innovation Initiative Consortium (QIIC) in Japan [58], and the Consortium organized around the Lomonosov Moscow State University in Russia (37). There are non-profit organizations with education and outreach purposes like QWorld [59], One Quantum [60], Q-munity [61], Feynman Foundation [62], and Quantum AI Foundation [63]. There are supporting organizations like Unitary Fund [64] and Quantum Open Source Foundation [65]. Finally, there are incubators and similar initiatives supported by local governments such as the Munich Quantum Valley [66] in Bavaria/Germany, the Quantum Algorithms Institute [67] in British Columbia/Canada, and Chicago Quantum Exchange [68] in Illinois/US. We believe that a landscaping study on the ecosystem comprising these organizations is also required as both the number of organizations, and their intertwinement with the rest of the ecosystem appears to be growing.

**Conclusion**

In this paper, we presented the gathered dataset of 441 companies from 42 countries that we identified as quantum start-ups, meaning that they were founded as companies mainly focusing on QT as their primary priority business. Following this, we answered some questions on temporal and geographical distribution of these start-ups, their categories, and patterns emerging from their analysis.

As a descriptive study, we hope this work can act as a good foundational work for future studies. The dataset can be improved by adding further information on companies like their size, valuation, investment amounts, and other aspects. Similarly, each field deserves more in-depth analysis of its dynamics, and some questions are raised here such as the effects of the emerging cohort of supporting companies, whether the start-up activity can actually be connected to following policy decisions (in the form of national initiatives or programs), and what are the path-dependencies that are currently being set which will limit the opportunity space of future development of the field. Start-ups are considered to be an important aspect of the emerging ecosystem of QT [69], and exploring such questions can lead to a better-informed understanding of further steps to be taken.

**Abbreviations**
EIC: European Innovation Council
PQC: Post-Quantum Cryptography
QT: Quantum Technologies
QED-C: Quantum Economic Development Consortium
QuIC: European Quantum Industry Consortium
QUTAC: Quantum Technology and Application Consortium
QKD: Quantum Key Distribution
SPAC: Special Purpose Acquisition Company

**Declarations**
*Ethical Approval and Consent to participate*
Not applicable.

*Availability of supporting data*
The datasets used and/or analysed during the current study are available from the corresponding author on reasonable request.

*Competing interests*
The authors declare that they have no competing interests.

*Funding*
Not applicable.

*Authors' contributions*
Z.C.S. constructed the dataset and wrote the main manuscript text. R.K. ran analyses and prepared the figures. A.U.A. guided the research, provided theoretical background, and edited the manuscript. All authors reviewed the manuscript.

*Acknowledgements*

We would like to thank Michel Kurek, Amanda Stein, and Kelvin Willoughby, Maninder Kaur, and Araceli Venegas-Gomez for insightful discussions and their feedback.

ZCS acknowledges support from the DAAD.